# Measurement and Analysis of Thermal Conductivity of Ceramic Particle Beds for Solar Thermal Energy Storage

Ka Man Chung[1]*, Jian Zeng[2]*, Sarath Reddy Adapa[2], Tianshi Feng[2], Malavika V. Bagepalli[3], Peter G. Loutzenhiser[3], Kevin J. Albrecht[4], Clifford K. Ho[4], Renkun Chen[#1,2]

[1]Material Science and Engineering Program, University of California, San Diego, La Jolla, California 92093, USA

[2]Department of Mechanical and Aerospace Engineering, University of California, San Diego, La Jolla, California 92093, USA

[3] George W. Woodruff School of Mechanical Engineering, Georgia Institute of Technology, Atlanta, Georgia 30332-0405, USA

[4] Concentrating Solar Technologies Department, Sandia National Laboratories, 1515 Eubank Blvd. SE, Albuquerque, NM 87123, USA

[#]Corresponding author: rkchen@ucsd.edu

*These authors contributed equally

**Abstract**

A systematic study was performed to measure the effective thermal conductivity of ceramic particle beds, a promising heat transfer and thermal energy storage media for concentrating solar power (CSP). The thermal conductivity of the ceramic particle beds was measured using a transient hot-wire (THW) method within a temperature range of room temperature to 700 °C, the target operating temperature of the next-generation CSP systems. Two different types of ceramic particles were examined: (1) CARBOBEAD HSP 40/70 and (2) CARBOBEAD CP 40/100 with the average particle sizes of ~ 400 μm and ~280 μm, respectively, and thermal conductivities ranging from ~0.25 W m$^{-1}$ K$^{-1}$ to ~0.50 W m$^{-1}$ K$^{-1}$ from 20 °C to 700 °C in both air and N$_2$ gas. The gaseous pressure dependence of the thermal conductivity of the ceramic particle beds was also studied in the $N_2$ environment to differentiate the contributions from gas conduction, solid conduction, and radiation. Calculations using the Zehner, Bauer, and Schlünder (ZBS) model showed good agreements with the measurements. Based on the model, it is concluded that the effective thermal conductivity of the packed particle beds is dominated by the gas conduction while the solid conduction and radiation contributes to about 20% of the effective thermal conductivity at high temperature.

## 1. Introduction

Concentrating solar power (CSP) applications coupled to thermal energy storage have continuously gained attention in the recent years due to their capability to store the thermal energy at low cost for on-demand electricity generation [1, 2]. One of the main objectives of the R&D efforts in CSP is to increase the efficiency by elevating the operating temperatures to above 700 ºC from the current limit of ~ 550 ºC [1-3]. A key component of the next-generation CSP system is the heat transfer medium. Conventional heat transfer media such as oil and molten nitrate salts have stable temperature limits of less than 400 ºC [4] and 600 ºC [5] respectively; molten chloride salts, on the other hand, can push the temperature up to ~800 ºC and are considered a candidate heat transfer fluid for the next-generation CSP. However, there are challenges associated with the chloride salts due to the corrosiveness to many containment materials [4, 5]. Solid particles are another type of promising heat transfer media for next-generation CSP systems as they can reliably operate at elevated temperatures with minimal safety concerns [6-8]. Studies of using falling ceramic particle beds in a CSP system have been ongoing since the 1980s [9] with several on-sun tests of particle-based CSP plants coupled to falling particle receiver [3, 10, 11]. Several of these ceramic particles have high absorptance in the solar spectrum for direct absorption of solar energy.

Upon heating with the solar energy, the ceramic particles can transfer the heat to another heat transfer fluid such as supercritical $CO_2$ ($sCO_2$) for power generation through a heat exchanger. Another important application of solid particles is thermal energy storage where heat is transferred from another heat transfer fluid (e.g., $sCO_2$) to particles through a heat exchanger and then stored in the high-temperature particles. Therefore, particle heat exchangers, such as moving packed particle beds-to-$sCO_2$ heat exchangers [12-15], are important for both thermal energy collection and storage.

Thermal conductivity of the particles as a heat transfer medium is critical for the performance of the particle heat exchangers. Among different types of ceramic particles, CARBOBEAD HSP 40/70 and CARBOBEAD CP 40/100 (also known as ACCUCAST ID 50) are two promising candidates being considered for the next-generation CSP systems. They are inexpensive, durable, and can maintain high solar weighted absorptance at high temperatures [2, 3, 12, 16]. While the optical properties, thermal stability, and flowing properties of this class of ceramic particles have been studied previously [2, 7, 16-18], a systematic study of their thermal conductivity is not yet available. Baumann and Zunft previously reported the effective thermal conductivity of various granular materials, such as quartz sand, sintered bauxite and corundum, with sizes ranging from 0.56 mm to 2.02 mm under atmospheric pressure and in the temperature range of room temperature to 800 ºC [19]. However, these particles are generally larger than CARBOBEAD HSP 40/70 and CP 40/100 , which have the sizes of ~ 400 μm and ~280 μm respectively. Furthermore, Baumann and Zunft did not discuss the roles of each heat transfer modes in the packed particle beds, thus making it difficult to predict or optimize the thermal conductivity. One of the key challenges of measuring the overall effective thermal conductivity of packed particle beds and differentiating the thermal conductivity due to different heat transfer modes is that different heat transfer pathways are convoluted in a single thermal conductivity measurement. Well controlled experiments must be made to delineate the heat transfer mechanisms of packed particle beds.

Due to the lack of experimental results, efforts over the last decades have been made to predict the thermal conductivity of the ceramic particle beds using analytical models [20-33] accounting gas and solid conduction and radiation contribution. These models predict the effective thermal conductivity ($k_{eff}$) of packed particle beds with sizes, shapes, porosities, packing structures and different materials as input parameters. The Zehner, Bauer, and Schlünder (ZBS) model is one of the widely used models, which is constructed based on a representative geometry method [15, 31-33] assuming particles in contact within a cylindrical unit cell. Although the ZBS model has been used to predict the thermal conductivity of HSP 40/70 and CP 40/100 [15], it is yet to be validated with experimental measurements. In addition, due to the uncertainty in the fitting parameters, the contributions of different heat transfer modes are still not accurately quantified.

In this study, we conduct a systematic and parametric study of the effective thermal conductivity of the packed particle beds of HSP 40/70 and CP 40/100 up to 700 ℃, the targeted operational temperature of the next-generation CSP plants. A transient hot wire (THW) setup is developed to systematically measure the thermal conductivity of these particles as a function of gaseous pressure and temperature, therefore, delineating the contributions of gas conduction, solid conduction, and thermal radiation. We also study the effect of aging and heating treatment on the thermal conductivity due to relevance towards the practical applications. The experimental results at different pressures and temperatures are fitted with the ZBS model to quantify the contribution from each heat transfer mode, with most of the fitting parameters (porosity, emissivity, bulk solid thermal conductivity) obtained from experiments. This study will provide useful data for the design of particle heat exchanger and will also lead to a better understanding of heat transfer mechanisms in ceramic particle beds.

## 2. Experimental Section

Two types of ceramic particle beds were investigated: (1) fresh CARBOBEAD CP 40/100 and (2) fresh CARBOBEAD HSP 40/70 (Carbo Ceramics, Inc.). Aged HSP 40/70 particles were also tested to study the possible degradation effect. These aged particles have undergone multiple runs of testing in falling particle towers and heat exchangers at Sandia National Laboratory [7, 10, 12], thus having been exposed to high solar flux, attribution, and thermal cycles. The general physical properties and material characterization of the ceramic particles were obtained using the following methods. The solid densities of the particles ($\rho_s$) were measured based on the Archimedes' principle, while the apparent density of the particle bed ($\rho_b$) was determined from the measured weight of the particle bed in the THW holder and its occupied volume in the cavity. The average particle size and the surface morphology were determined by the images taken from FEI Quanta 250 scanning electron microscope (SEM). Images of 100 particles were analyzed to determine the average diameter $d_p$, average circularity and roundness using the software imageJ [34-36]. In the software, the boundary of a particle in a SEM image was selected. The distances between the selected boundaries of the particles were then averaged out to obtain the average diameter $d_p$. Three additional parameters were used to quantify the morphology of the particles: circularity ($C$), aspect ratio ($A_r$), and roundness ($R$). $C$ of a particle is defined as $C = \frac{4\pi A}{P^2}$ where $A$ is the cross-sectional area (as viewed in the SEM images) and $P$ is the perimeter of the particle. $A_r$ is defined as the ratio between the major axis ($a$) and the minor axis ($b$) of a particle. $R$ of a particle is defined as $R = \frac{4A}{\pi a^2}$.

The chemical compositions of the particles were determined by the Energy Dispersive X-Ray Spectroscopy (EDS) in the SEM. X-ray diffraction (XRD) measurement was performed using the Rigaku Smartlab X-ray diffractometer to obtain the X-ray pattern of each particle for chemical identification.

To measure the effective thermal conductivity ($k_{eff}$) of particle beds at different temperatures and gas pressures, we have designed and fabricated a high-temperature THW apparatus capable of measuring $k_{eff}$ within a wide temperature range (room temperature to about 700 °C) and at different gas pressure levels (1 to 760 Torr). The THW method is a conventional thermal conductivity measurement method for various substances (*i.e.*, liquids, compressed gases, and powders [37-42]). **Figure 1a and 1b** show the schematic and image of the high-temperature THW setup, respectively, while **Figure 1c** shows a photograph of the THW test section. The THW test section was made of mica (McMaster-Carr) capable of withstanding the maximum operating temperature of ~798 °C. A cavity with a dimension of roughly 0.97 (width) × 1.50 (depth) × 11.6 (length) cm$^3$ was machined to contain the particles for the measurement. A Pt wire (Goodfellow Corporation, PT0005114) of 25.0 μm in diameter was embedded in the middle of the cavity, serving as the hot wire in the THW measurement as well as a resistive thermometer of the particles. The temperature of the particles was additionally monitored by a K-type thermocouple inserted directly into the particle bed. The thermocouple was connected to a thermocouple monitor (Stanford Research Systems SR630). Four Cu wires were fixed by set-screw connectors (Kurt J. Lesker Company, FTASSC050) at the end of the mica holder, enabling good electrical contacts to the Pt wire. This four-wire configuration is essential to eliminate the contact resistance when measuring the resistance of the Pt hot wire, as shown in the inset in **Figure 1a**.

To measure $k_{eff}$ at different temperatures and gaseous pressures, the THW test section was placed inside a tube furnace (MTI GSL-1100X quartz tube furnace) with a maximum continuous operation temperature of 1000 ºC. The furnace is connected to a gas cylinder in the upstream and a mechanical pump in the downstream. The four Cu wires were connected to pins of a vacuum feedthrough. The external terminals of the feedthrough were connected to a constant programmable current source (Keithley 220) and a digital multimeter (Hewlett-Packard 34401A, 6½ Digit), which are controlled by a computer through a GPIB interface. When the gas was introduced to the quartz tube, the gas flow rate was monitored and controlled by a mass flow controller (MKS Instruments 179C) that was connected to a 4-channel flow controller power supply readout (MKS Instruments 247D). On the downstream side, a pressure transducer (MKS Baratron Type 127A) and an exhaust throttle valve (MKS Instruments 253B-26373) were installed in series to monitor and regulate the gaseous pressure inside the tube furnace.

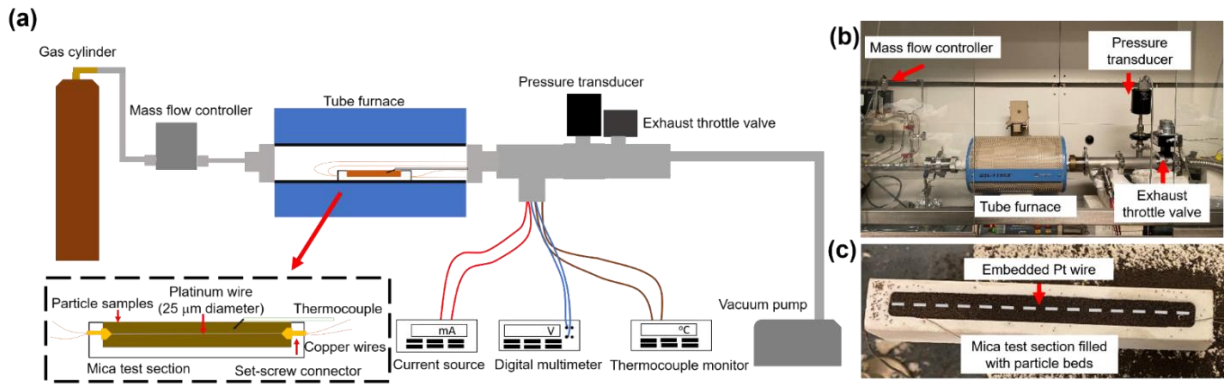

**Figure 1. (a)** Schematic of the THW set-up; **(b)** photograph of the THW measurement set-up; and **(c)** photograph of THW test section with particle samples filled inside.

In the THW measurement, with the assumption that a line heat source is embedded in an infinite particle bed, the effective thermal conductivity of the particles is given by the following equation [41-45]:

$$k_{eff} = \frac{q'/4\pi}{d(\Delta T)/d(\ln t)} \quad (1)$$

where $\Delta T$ is the temperature rise of the Pt wire induced by applying an electrical current $I$ to it for a duration of time $(t)$, $q'$ is the heat rate per unit length of the Pt wire: $q' = I^2 R/L$, where $L$ is the length of the Pt wire, and $R$ is the resistance of the Pt wire at the measurement temperature $T$. In our measurement, $I$ usually ranged from 50-100 mA to achieve a sufficiently large $\Delta T$ in the particle samples. $\Delta T$ was determined from the measured resistance change ($\Delta R$) of the Pt wire upon heating with the applied current:

$$\Delta T = \Delta R/(R_{ref}\beta) \quad (2)$$

where $R_{ref}$ is the Pt resistance at room temperature (23 °C) and the temperature coefficient of resistance $\beta$ is evaluated at the measurement temperature $T$:

$$\beta = \frac{1}{\rho_{ref}} \frac{d\rho(T)}{dT} \quad (3)$$

where $\rho_{ref}$ is the resistivity of Pt at room temperature and the temperature-dependent resistivity $\rho(T)$ of solid Pt was obtained from the known polynomial equation for the temperature range of 100 to 2041.3 K [46]. Additionally, to ensure that the thermometry based on the literature value of $\beta$ for Pt is valid, we measured the temperature of the particle bed using a K-type thermocouple probe inserted in the bed. The temperature obtained from the thermocouple was consistent with that extracted from the measured resistance of the Pt wire using the $\beta$ value in Eq. 3 (see **Figure S1** in Supplementary Materials).

To ensure that the thermal conductivity measurement is only sensitive to the thermophysical properties of the packed particles filled within the cavity of the test section, the heating duration and measurement time $t$ was chosen such that the thermal penetration depth $L_p = \sqrt{4\alpha t}$, where $\alpha$ is the thermal diffusivity of the particles, is larger than the diameter of the Pt wire while smaller than half of the width and height of the cavity of the mica holder.

To validate the THW setup, deionized (DI) water and two oil-based heat transfer fluids, Dowtherm A oil and XCLETHERM 600, were used as the reference materials for calibration. DI water and Dowtherm A were measured at room temperature, and XCELTHERM 600 was measured from room temperature to 225 ºC.

To model $k_{eff}$ of CARBO ceramic particle beds, the intrinsic thermal conductivity of the solid ($k_s$) is needed. Solid pellets were obtained by hot pressing the CARBOBEAD CP particles into a nearly fully dense form (pellet density $\rho_P = 3041$ kg m$^{-3}$). The pellet was machined to a circular disk shape of 12.5 mm in diameter and 3.02 mm in thickness. The thermal diffusivity $\alpha_s$ of the solid pellet was determined using a laser flash analyzer (LFA, NETZSCH HyperFlash 467 HT) from room temperature to 800 ºC. The specific heat capacity $c_p$ of the CARBO ceramic particles was extracted from Ref. [16]. The thermal conductivity of the CARBO ceramic solid pellet ($k_P$) was then determined by:

$$k_P = \alpha_s \rho_P c_p \tag{4}$$

Subsequently, the thermal conductivity of the solid particles can be obtained as:

$$k_s = \frac{\rho_s}{\rho_P} k_P \tag{5}$$

## 3. Results and Discussion

### 3.1. Calibration of the THW setup

The calibration of the THW setup was done by measuring three standard fluids: (1) DI water, (2) Dowtherm A oil, and (3) XCELTHERM 600. As shown in **Figure 2(a)**, the measured thermal conductivity of DI water, Dowtherm A oil and XCELTHERM 600 at room temperature are 0.6093±0.0084 Wm$^{-1}$K$^{-1}$ (literature value: 0.6009 Wm$^{-1}$K$^{-1}$ [47, 48]), 0.1382±0.0006 (literature

value: 0.1380 Wm$^{-1}$K$^{-1}$ [47]), and 0.1366±0.0006(literature value: 0.1359 Wm$^{-1}$K$^{-1}$ [49]). For XCELTHERM 600, the thermal conductivity was also measured from 20 °C to 225 °C and was compared to the recommended values, as shown in **Figure 2**. The standard deviation of each data point shown on the figure is generally less than 0.70%. The deviation of the experimental results from the literature values is < 0.50%. Therefore, the THW setup can be considered well calibrated.

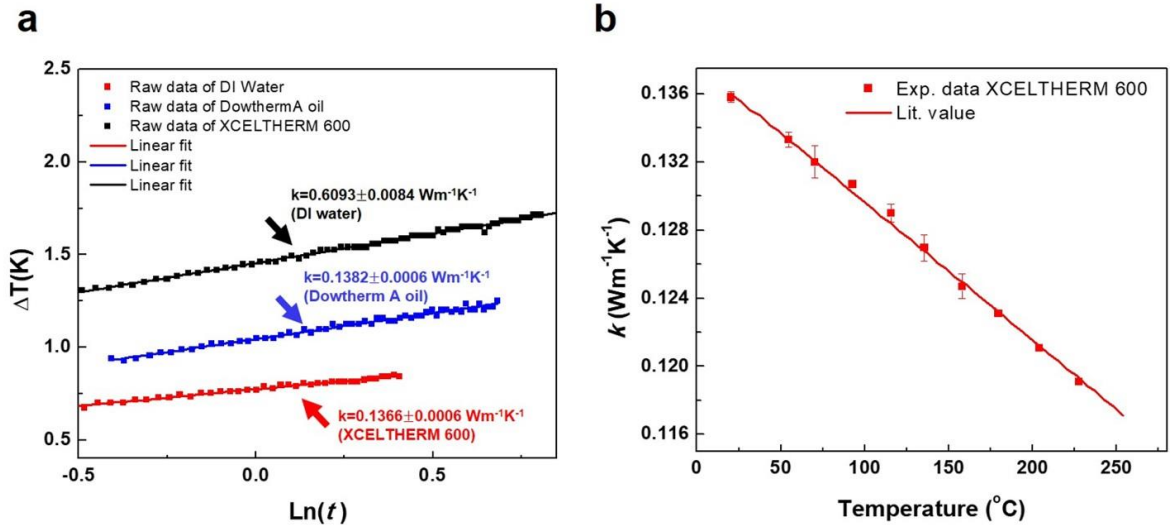

**Figure 2 (a)** Plot of $\Delta T$ versus Ln ($t$) for DI water, Dowtherm A oil, and XCELTHERM 600 oil at room temperature. The extracted thermal conductivity values are shown next to the plots of each fluid. **(b)** Calibration results of the THW measurement using XCELTHERM 600 oil from 20 °C to 225 °C.

### 3.2. Characterization of particles

**Table 1** summarizes the physical and geometrical properties of the ceramic particles. From **Table 1**, the density of fully dense ceramic particle is around 3500 kg m$^{-3}$, which is similar to the density of pure Aluminum Oxide (Al$_2$O$_3$) (range from 3000 to 3980 kg m$^{-3}$) [49]. This is consistent with the compositional analysis result showing that Al$_2$O$_3$ is one of the major constituents in the ceramic particles, with small amounts of SiO$_2$, Fe$_2$O$_3$, and TiO$_2$, etc (see **Figure S2** and **Figure S3** in Supplementary Materials).

**Table 1** Summary of the physical and geometric properties of the measured CARBOBEAD ceramic particles

| Particle sample | Solid density of particles $\rho_s$ (kg m$^{-3}$) | Apparent density of particle bed $\rho_b$ (kg m$^{-3}$) | Porosity $\varepsilon$ $(1-\rho_b/\rho_s) \times 100\%$ | Average particle diameter $d_p$ (μm) | Average circularity $C$ | Average roundness $R$ | Average aspect ratio $A_r$ |
|---|---|---|---|---|---|---|---|
| Fresh CARBOBEAD CP 40/100 | 3530 | 2015 | 42.9 | 275 | 0.673 | 0.823 | 1.270 |
| Fresh CARBOBEAD HSP 40/70 | 3480 | 2164 | 38.7 | 404 | 0.572 | 0.770 | 1.548 |
| Aged CARBOBEAD HSP 40/70 | 3480 | 2083 | 41.0 | 403 | 0.508 | 0.722 | 1.600 |

**Figure 3** shows the SEM images of each particle specimen. From these images, we extracted $d_p, C, R, A_r$ as described in the *Experimental Section*. The results are also shown in **Table 1**. Fresh and aged HSP 40/70 particles have similar average particle diameters of $d_p \sim 400$ μm, while the fresh CP 40/100 has $d_p = 275$ μm. The shape of the HSP 40/70 and CP 40/100 particles are not strictly spherical. Instead, they exhibit irregular and elongated shapes. In addition, the shapes of individual particles vary within the same particle batch. The circularity $C$ of a particle is a measure of its similarity to a perfect circle [34, 36]. When $C = 1$, it denotes that the particle is perfectly circular. The aspect ratio $A_r$, on the other hand, is the ratio between the major *a,* and minor axes *b* of a particle, which is a measure of the degree of elongation of a particle, [34, 36]. For a perfect sphere, $A_r = 1$. The roundness $R$ of a particle measures the curvature of the edge of the particle and provides further understanding on the shape of the particle at small scales [34, 51]. The circularity of CP 40/100, HSP 40/70, and aged HSP 40/70 are 0.673, 0.572 and 0.508, respectively, revealing the asymmetric geometry of the particles. The average roundness of the particles ranges from ~0.7 to 0.8, indicating that the particles are rounded in general. The aspect ratios of the three particle samples are between ~1.3 to ~1.6 indicating that the particles are elongated. **Figure 3d, e and f** show the SEM images of the surface of the particle samples. It is evident that the particle surfaces are rough. For the aged HSP particles, most of them are similar to the fresh HSP particles. However, some of them show cracks and fractures on the surface, as shown in **Figure 3f**, which may be caused by the collisions occurred during the free falls in the falling particle tower.

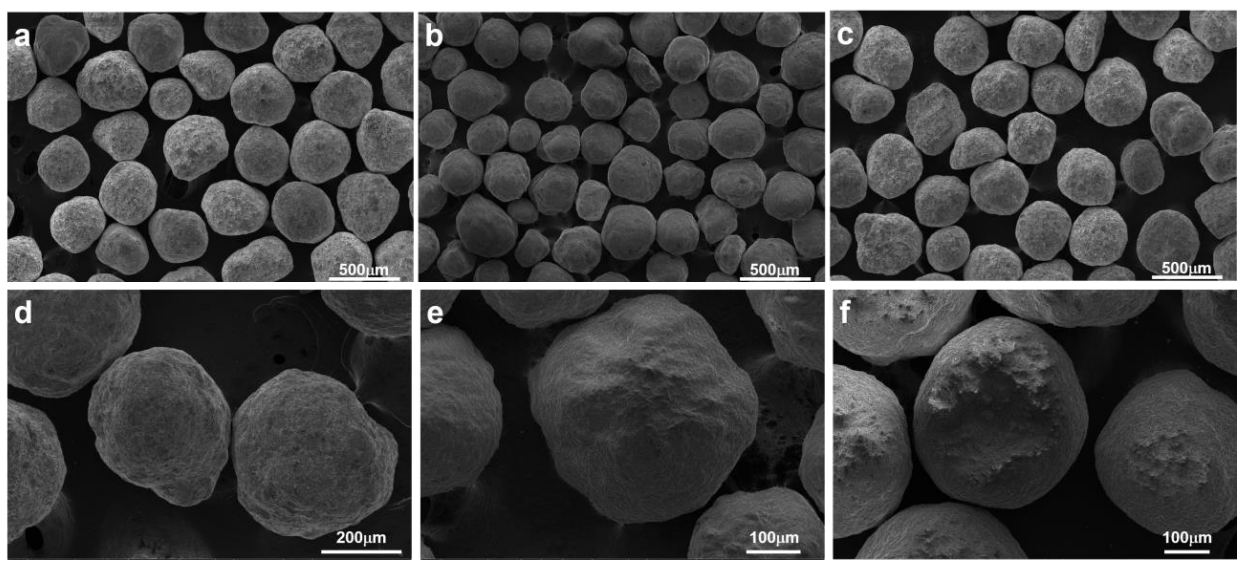

**Figure 3.** SEM images of **(a,d)** fresh CARBOBEAD HSP 40/70; **(b,e)** fresh CARBOBEAD CP 40/100; and **(c,f)** aged CARBOBEAD HSP 40/70

### 3.3. Effective thermal conductivity under air and $N_2$ gas environments

The effective thermal conductivity of HSP 40/70 and CP 40/100 under air and $N_2$ gas is shown in **Figure 4a**. The error bars in the figure include both the standard deviations of measurements of at least three samples for each type of particles and the systematic error obtained from the standard error propagation analysis (see Supplementary Materials Note 1). Under either air or $N_2$ gaseous environment, the HSP 40/70 particles with larger $d_p$ possess higher thermal conductivity than CP 40/100. Both HSP 40/70 and CP 40/100 particles show similar $k_{eff}$ between air and N₂ gas environments at 760 Torr. This is reasonable because the bulk gas thermal conductivity ($k_g$) of air and N₂ are similar ($k_g$ = 0.00263 Wm⁻¹K⁻¹ for air [48] and $k_g$ = 0.00259 Wm⁻¹K⁻¹ for N₂ gas at $T$ = 300 K [48]). Therefore, the gas conduction contribution to the overall effective thermal conductivity of the ceramic particles is similar when the gas type is changed from air to N₂ gas.

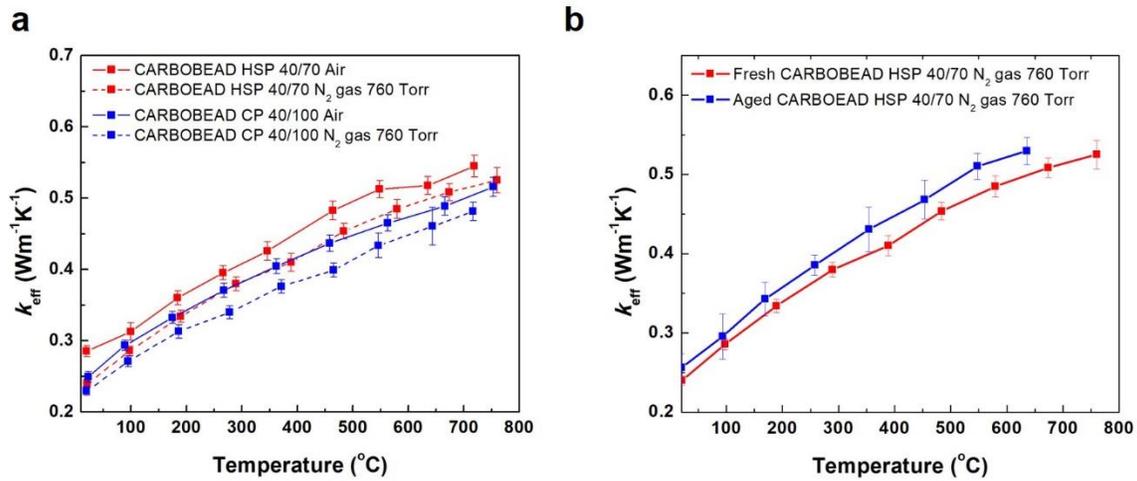

**Figure 4 (a)** Effective thermal conductivity as a function of temperature for CARBOBEAD HSP 40/70 and CARBOBEAD CP 40/100 in air and N₂ gas at 760 Torr; **(b)** Effective thermal conductivity as a function of temperature for fresh and aged HSP 40/70 under N₂ gas at 760 Torr.

**Figure 4b** shows $k_{eff}$ of fresh and aged HSP 40/70 particles under 760 Torr N₂ as a function of temperature. There is no substantial difference in the thermal conductivity between the fresh and aged particles. From the material characterization, including SEM coupled with EDS and XRD (see **Figure S2** and **Figure S3** in Supplementary Materials), there is only minor difference in the particle size and shape between the fresh and aged HSP particles. Despite the cracks and fractures on the surfaces of some aged CARBO particles, which may have been caused by the collisions of particles during the free falling in the receiver under the on-sun tests, they do not change the general particle morphology significantly according to the particle shape analysis shown in **Table 1**. Therefore, the aging condition of the HSP particles has minimal effect on the heat transfer performance of the particle beds.

## 4. Pressure dependence of effective thermal conductivity

The gaseous pressure dependence of $k_{eff}$ of the packed particles are shown in **Figure 5. Figure 5a and b** show $k_{eff}$ of the particles as a function of temperature under $N_2$ gaseous pressure from 760 Torr to 1 Torr. $k_{eff}$ of the particle beds drops substantially when the gaseous pressure decreases from 760 Torr to 1 Torr. At 1 Torr, $k_{eff}$ shows a nearly constant value within the measured temperature range. The pressure dependent $k_{eff}$ reveals that solid-to-gas and gas conduction play a dominant role in the entire temperature range. Solid-to-solid conduction, on the other hand, only contributes slightly to the overall thermal conductivity.

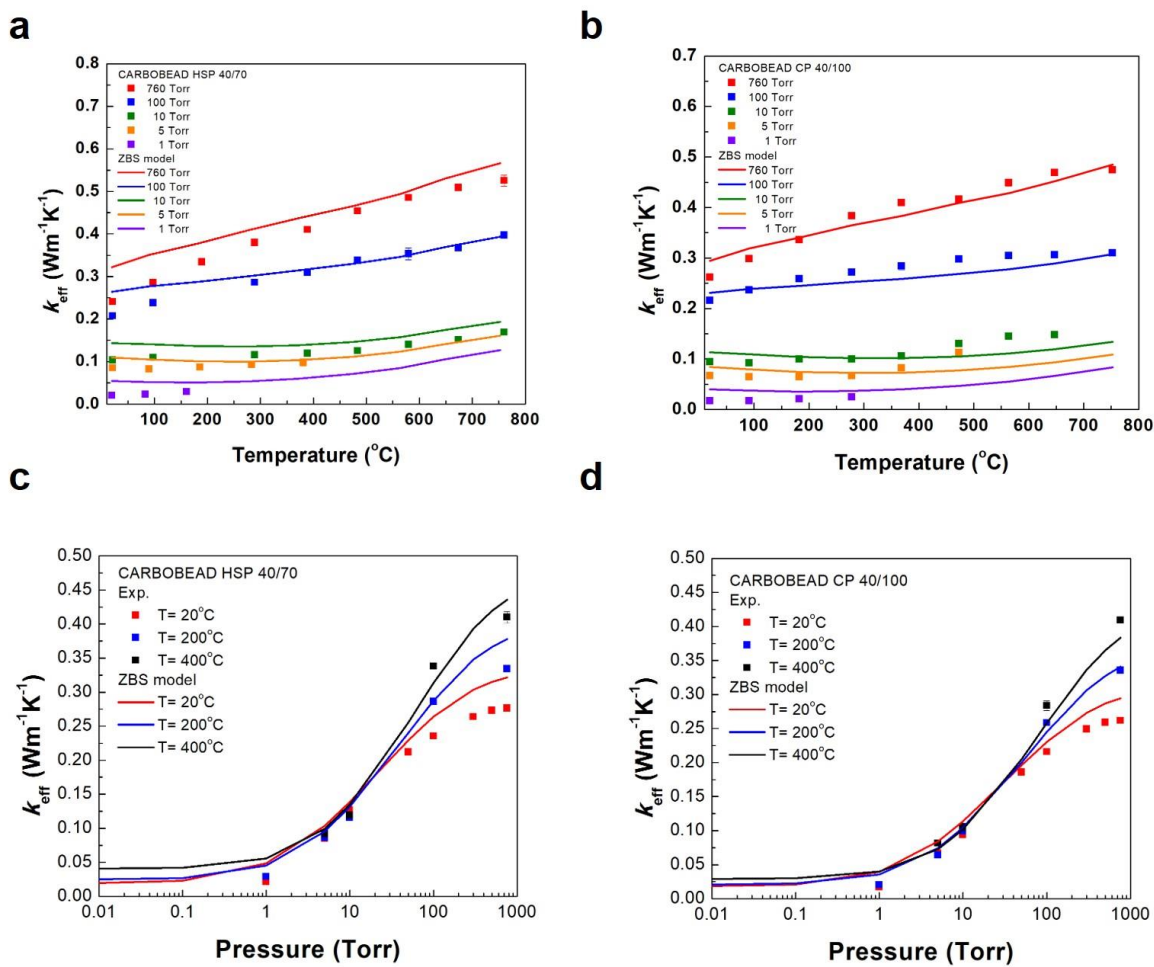

**Figure 5**. Effective thermal conductivity as a function of temperature for **(a)** CARBOBEAD HSP 40/70 and **(b)** CARBOBEAD CP 40/100 under different $N_2$ pressures, 760, 100, 10, 5 and 1 Torr; effective thermal conductivity ($k_{eff}$) as a function of gas pressure for **(c)** CARBOBEAD HSP 40/70 and **(d)** CARBOBEAD CP 40/100 at different temperatures, including 20 ºC, 200 ºC and 400 ºC.

**Figures 5c and d** show $k_{eff}$ as a function of gaseous pressure at 20, 200, and 400 ºC. The figures again show decreasing trends of $k$ with reducing gaseous pressure at all the temperatures for both types of the particles. At the high gas pressure limit (760 Torr), $k_{eff}$ increases with the temperature, principally due to the increasing bulk gas thermal conductivity at high temperatures. At low gas pressure limit (1, 5, and 10 Torr), $k_{eff}$ is greatly reduced with little to no temperature dependence below 400 °C, as evidenced by the overlapped data below 10 torr in **Figures 5c and 5d**. This is because at low gas pressure and low-to-intermediate temperatures, the gas contribution is diminished and thermal radiation contribution is very small, and thus, $k_{eff}$ is only a function of the solid conduction, which has weak temperature dependence.

## 5. Solid thermal conductivity ($k_s$) of CARBO ceramic pellet

To model $k_{eff}$ of the CARBO ceramic particle beds, the thermal conductivity of the corresponding solid material in the fully dense bulk form ($k_s$) is needed. **Figure 6** show the thermal diffusivity ($\alpha_s$) measured using the LFA, and the measured thermal conductivity of the CARBO ceramic pellet ($k_P$) determined from Eq. (4). $k_P$ of the pellet deceases from ~6.50 Wm$^{-1}$K$^{-1}$ to ~4.50 Wm$^{-1}$K$^{-1}$ as the temperature increases from 20 °C to 800 °C. The deceasing trend is consistent with the typical thermal conductivity of bulk solids with similar compositions (*i.e.*, Al$_2$O$_3$ and SiO$_2$). $k_P$ is then converted to $k_s$ by considering the density difference between the pellet and solid particles (Eq. (5)).

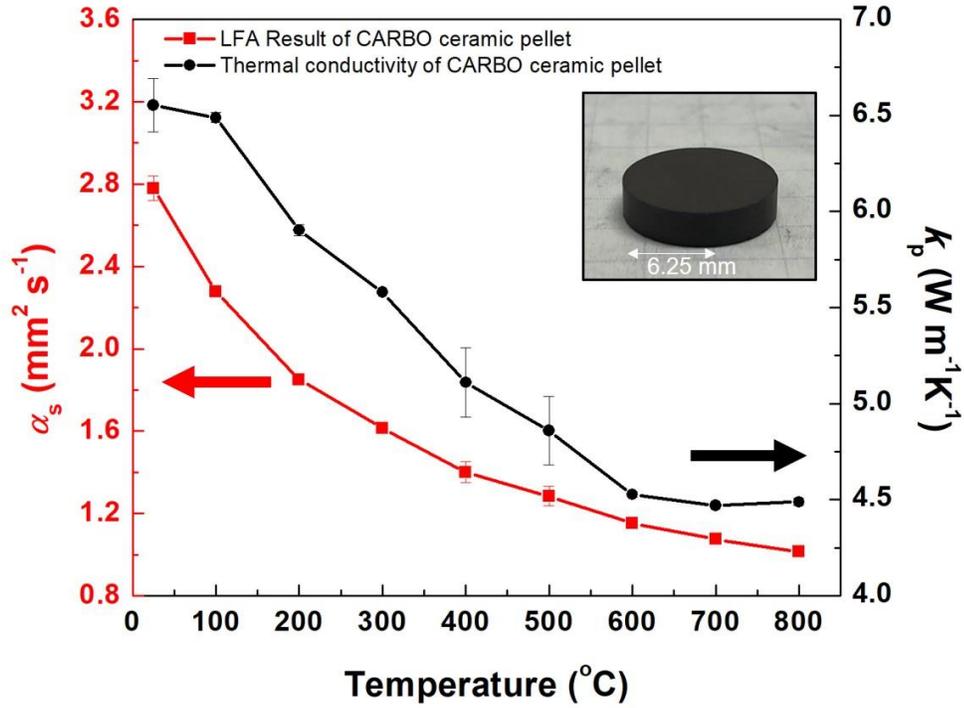

**Figure 6.** Measured thermal diffusivity ($\alpha_s$) and thermal conductivity ($k_P$) of the CARBO ceramic solid pellet based on the LFA measurement. Inset: photograph of the solid pellet made from CARBOBEAD CP particles via hot pressing.

## 6. Analysis using the ZBS Model

The ZBS model was employed to fit the experimental results of the effective thermal conductivity of HSP 40/70 and CP 40/100, as a function of temperature and gaseous pressure. In the ZBS model, $k_{eff}$ is given by [31]:

$$\frac{k_{eff}}{k_f} = \left[1 - (1-\varepsilon)^{\frac{1}{2}}\right]\varepsilon\left[\left(\varepsilon - 1 + \frac{1}{\lambda_G}\right)^{-1} + \lambda_r\right] + (1-\varepsilon)^{\frac{1}{2}}[\varphi\lambda + (1-\varphi)\lambda_c] \quad (6)$$

where $k_f$ is the thermal conductivity of the bulk gas at the corresponding pressure and temperature, $\varepsilon$ is the porosity of the packed particle beds, and $\varphi$ is the empirical contact fraction of the solid particle. The dimensionless parameters, $\lambda_G, \lambda_r$, and $\lambda$, correspond to the heat transfer processes through gas conduction, radiation, and solid conduction at the contact region, respectively. For the gas conduction,

$$\lambda_G = \frac{k_G}{k_f} = \left[1 + \left(\frac{l}{d_p}\right)\right]^{-1}, \tag{7}$$

where $k_G$ is the thermal conductivity of gas within the particle bed and $l$ is the modified mean free path of the gas molecules, which can be calculated at a given temperature and pressure by following [31]. For the radiation parameter,

$$\lambda_r = \frac{4\sigma}{\left(\frac{2}{\varepsilon_r}-1\right)} T^3 \frac{d_p}{k_f}, \tag{8}$$

where $\sigma$ and $\varepsilon_r$ are Stephan-Boltzmann constant and emissivity of the particles, respectively. For the solid conduction parameter,

$$\lambda = \frac{k_s}{k_f} \tag{9}$$

and

$$\lambda_c = \frac{2}{N}\left\{\frac{B(\lambda+\lambda_r-1)}{N^2\lambda_G\lambda} \ln\frac{\lambda+\lambda_r}{B[\lambda_G+(1-\lambda_G)(\lambda+\lambda_r)]} + \frac{B+1}{2B}\left[\frac{\lambda_r}{\lambda_G} - B\left(1+\frac{1-\lambda_G}{\lambda_G}\lambda_r\right) - \frac{B-1}{N\lambda_G}\right]\right\} \tag{10}$$

where $B = 1.364\left(\frac{1-\varepsilon}{\varepsilon}\right)^{1.055}$ is the deformation parameter and $N = \frac{1}{\lambda_G}\left(1 + \frac{\lambda_r - B\lambda_G}{\lambda}\right) - B\left(\frac{1}{\lambda_G} - 1\right)\left(1 + \frac{\lambda_r}{\lambda}\right)$

There are five parameters in the ZBS model: $d_p, \varepsilon, \varepsilon_r, k_s$ and $\varphi$. Among these, $d_p, \varepsilon, \varepsilon_r$, and $k_s$ are experimentally determined: $\varepsilon$ and $d_p$ are obtained from **Table 1**, $\varepsilon_r$ is extracted from **Ref. [52]** where $\varepsilon_r$ is 90.8% for HSP 40/70 and is 86.8% for CP 40/100, leaving $\varphi$ as the only fitting parameter, which was obtained by fitting the modeling results to the experimental results for both types of particles at all the test temperatures and gas pressures, resulting in $\varphi = 0.0031$. The results from the ZBS model are shown alongside with the experimental data in **Figure 5**. The ZBS model can well capture the effective thermal conductivity of both HSP40/70 and CP 40/100 packed particle beds as function of temperature and gaseous pressure. For example, the curves representing the ZBS model in **Figure 5a** and **5b** show substantial decrease in the thermal conductivity when the gas pressure changes from 760 Torr to 100, 10 and 1 Torr. From **Figure 5c and d**, the ZBS modeling results accurately predict the measured $k_{eff}$ of HSP 40/70 and CP 40/100 in most of the gas pressure range except at 1 Torr.

At 1 Torr, however, the ZBS model tends to overestimate the $k_{eff}$ compared to the measured values. The overestimation by the ZBS model at the lowest gas pressure levels is not surprising because at this gas pressure level, gas conduction is greatly diminished and the solid conduction becomes more important, which can be highly dependent on the exact packing structure of the particles. However, the ZBS model is based on a simple cubic arrangement of the particles, whereas in reality the particle arrangement is random and more complex. This is evident from the difference between the experimental porosity ($\varepsilon \sim 38.7 - 42.9\%$, **Table 1**) and that of a simple cubic structure ($\varepsilon$ is expected to be 48%). The lower porosity of the actual packing structure is possibly the cause of the overestimated thermal conductivity by the ZBS model at 1 Torr. The modeling of $k_{eff}$ of the packed particles at this gas pressure requires a more realistic treatment of the particle packing structure, which is not within the scope of this study and warrants further investigation. Nevertheless, at higher gaseous pressures, especially at 760 Torr at which the particles are expected to be used in CSP systems, the gas conduction is the dominant heat transfer model and the effect of the packing structure is less important. This explains the applicability of the ZBS model at higher gaseous pressure despite its simplicity.

With this applicability in mind, we use the ZBS model to understand the roles of different heat transfer modes in particle beds under normal operation pressure of 760 Torr. **Figure 7** shows the calculated effective thermal conductivity of the two types of the particles using the ZBS model and differentiates the contributions from solid conduction, gas conduction and thermal radiation.

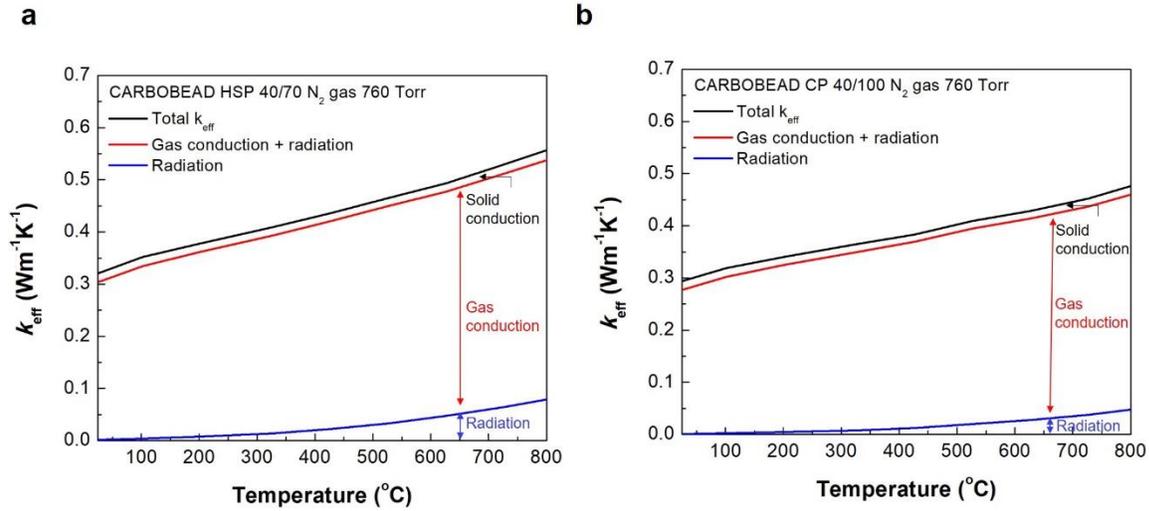

**Figure 7.** Plot of effective thermal conductivity calculated from the ZBS model showing individual contribution from solid conduction, gas conduction and thermal radiation: (a) CARBOBEAD HSP 40/70; and (b) CARBOHEAD CP 40/100, under $N_2$ gas environment at 760 Torr.

Based on **Figure 7**, gas conduction in these particle beds contributes to more than 80% of the total $k_{eff}$ in the whole temperature range; solid conduction from particle-to-particle, on the other hand, contributes to around 5% of the total effective thermal conductivity at room temperature, with its relative contribution decreasing due to the increasing total thermal conductivity. Regarding the thermal radiation contribution, it is not important until at elevated temperatures. At 800°C, around 15% (in HSP 40/70) and 11% (in CP 40/100) of the total $k_{eff}$ is contributed from the radiation heat transfer. The slightly larger radiation contribution in HSP 40/70 at 800 °C is due to the larger particle sizes, which lead to larger void sizes for longer photon propagation length. Nevertheless, in both types of particles, thermal radiation plays an insignificant role, due to the limited photon propagation length (optically thick) within the particle beds, which is caused by the high absorptance of the particles and the relatively small void size between the particles (comparable to the particle size, $d_p$). The analysis also suggests possible strategies to further enhance the thermal conductivity of the particle beds, *e.g.* by filling the voids in the particle beds with smaller particles to increase the gaseous thermal pathways bridging solid particles.

## 7. Conclusion

In this study, we measured the thermal conductivity of CARBOBEAD HSP 40/70 and CARBOBEAD CP 40/100 in $N_2$ gas and air environment under different gaseous pressures by developing a custom-made high-temperature THW setup. The effective thermal conductivity of both types of particles increases at higher temperature under both ambient air and $N_2$ at 760 Torr and decreases with lower $N_2$ gas pressure at the same temperature. The thermal conductivity is slightly higher for HSP 40/70, which has a larger average particle size compared to CP 40/100 (404 µm vs. 275 µm). We also found that the aged CARBO HSP particles show similar thermal conductivity as the fresh ones. The systematic temperature and gaseous pressure dependent thermal conductivity data are analyzed using the ZBS model that assumes a simple cubic particle packing structure, with mostly experimentally determined parameters (solid thermal conductivity, particle size, porosity, thermal emissivity) and only one single fitting parameter (solid contact fraction) for both particle types. Despite the simplicity, the ZBS model works well in explaining the experimental data obtained at various temperatures and gas pressures, except at the lowest pressure (1 Torr), where a realistic particle packing structure is needed to model the dominant particle-to-particle solid conduction effect. Through the combined experimental and modeling work, the relative contributions of gas conduction, particle-to-particle solid conduction, and thermal radiation to the particle bed thermal conductivity are elucidated. We found that gas conduction plays a dominant role within the entire temperature range under ambient pressure, with the remaining thermal conductivity contributed mainly by either solid conduction around room temperature or thermal radiation at high temperature. It is expected that experimental data and the analysis presented in this work can shed light on the heat transfer mechanisms in packed particle beds and could guide the design of particle-based heat exchangers for next-generation CSP plants.

**Acknowledgements**

This material is based upon work supported by the U.S. Department of Energy's Office of Energy Efficiency and Renewable Energy (EERE) under Solar Energy Technologies Office (SETO) Agreement Number DE-EE0008379. The views expressed herein do not necessarily represent the views of the U.S. Department of Energy or the United States Government.